\documentclass[aps,amssymb,12pt,showpacs,superscriptaddress,secnumarabic]{revtex4}
\usepackage{graphicx,t1enc}

\begin{document}
\date{\today}
\title{Bistability and Hysteresis in the Sliding Friction of a Dimer}

\author{S. Gonçalves} \email{sgonc@if.ufrgs.br}
\affiliation{Consortium of the Americas for Interdisciplinary Science and\\
Department of Physics and Astronomy, University of New Mexico, Albuquerque,
New Mexico 87131}
\affiliation{Instituto de Física, Universidade Federal do Rio
Grande do Sul, Caixa Postal 15051, 90501-970 Porto Alegre RS, Brazil}

\author{C. Fusco} \email{C.Fusco@science.ru.nl}
\affiliation{Consortium of the Americas for Interdisciplinary Science and\\
Department of Physics and Astronomy, University of New Mexico, Albuquerque,
New Mexico 87131}
\affiliation{Solid State Theory, Institute for Molecules and Materials,
Radboud University Nijmegen,\\
Toernooiveld 1, 6525 ED Nijmegen, The Netherlands}

\author{A. R. Bishop} \email{arb@lanl.gov}
\affiliation{Theoretical Division and Center for Nonlinear Studies,
Los Alamos National Laboratory, Los Alamos, New Mexico 87545}

\author{V. M. Kenkre} \email{kenkre@unm.edu}
\affiliation{Consortium of the Americas for Interdisciplinary Science and\\
Department of Physics and Astronomy, University of New Mexico, Albuquerque,
New Mexico 87131}

\begin{abstract} The sliding friction of a dimer moving over a periodic
substrate and subjected to an external force is studied in the steady state for
arbitrary temperatures within a one-dimensional model.  Nonlinear phenomena that
emerge include dynamic bistability and hysteresis, and can be related to earlier
observations for extended systems such as the Frenkel-Kontorova model. Several
observed features can be satisfactorily explained in terms of the resonance of a
driven-damped nonlinear oscillator.  Increasing temperature tends to lower the
resonant peak and wash out the hysteresis.
\end{abstract}

\pacs{81.40.Pq,46.55.+d}

\maketitle
\section{Introduction} The friction experienced by atoms, small molecules, and
adlayers moving over substrates is an active topic of current
research~\cite{Gnecco}.  One reason is the desire to understand, at a
fundamental level, the origin of friction. The other reason is the wish to
acquire expertise in developing nano-devices like nano-motors, nano-wires, and
nano-probes.  Besides, sliding friction is related to other atomistic processes
at surfaces, such as diffusion of atoms and
molecules~\cite{Ala,Vega,Braun,Romero}, and motion of long
chains~\cite{Strunz98,Strunz98b,Braun97,Paliy,Braun03} over periodic substrates.  That
microscopic sliding friction can exhibit nonlinear behavior depending on the
sliding regime has been shown in the recent theoretical literature.  Strunz and
Elmer~\cite{Strunz98} have studied in detail the nonlinear friction of the
Frenkel-Kontorova model and identified the origins of such friction as being the
resonance of the sliding velocity with the internal vibration modes of the
chain, and the formation of kinks.  The latter phenomenon was also studied by
Braun et al.~\cite{Braun97,Braun03}.  More recently, Fusco and Fasolino~\cite{Fusco03a,
Fusco03b} have identified the same resonance phenomenon in the friction of a
smaller object, a dimer moving over a periodic substrate.  Gonçalves et
al.~\cite{Goncalves04} have analyzed a closely related system in the relaxation
regime, i.e. in the absence of external forces, and have given a simple physical
explanation of the observed results. Persson~\cite{Persson93} has predicted that
a friction force proportional to $v^{-3}$ is to be expected (in addition to the
linear one) in the large-force regime for a sliding system of any size, ranging
from a single atom to an infinite chain.  In the present paper, we restrict our
study to the steady state friction of a dimer, but investigate all force
regimes. We identify several separate regimes exhibiting resonance, bistability,
and hysteresis. We provide a simple explanation for these phenomena in terms of
the resonance of a driven-damped oscillator, and comment on how they depend on
the substrate corrugation, damping and temperature.

\section{Model and Simulation results}
At zero temperature, the equations of motion for the two particles constituting
the dimer sliding over a periodic potential, in the presence of external force
$F$, are~\cite{Fusco03b,Goncalves04}:
\begin{eqnarray}
m\ddot x_{1} + m\gamma\dot x_{1} - k(x_{2} - x_{1} - a) =
\frac{2\pi u}{b}\sin\left(\frac{2\pi x_{1}}{b}\right) + F \nonumber \\
m\ddot x_{2} + m\gamma\dot x_{2} + k(x_{2} - x_{1} - a) =
\frac{2\pi u}{b}\sin\left(\frac{2\pi x_{2}}{b}\right) + F,
\label{x12}
\end{eqnarray}
where $x_{1,2}$ are the coordinates of the two particles each of mass $m$,
and $k$, $a$, $b$, $u$ are, respectively, the spring constant, equilibrium
length of the dimer, wavelength of the substrate potential and half the
amplitude of the potential. We integrate these coupled equations numerically
using the algorithm of Verlet modified to allow for velocity-dependent
forces~\cite{Goncalves04}.  The underlying characteristic physical quantities in
this system are the equilibrium dimer length $a$, the substrate wavelength $b$,
the free dimer characteristic time $1/\omega_0=\sqrt{m/2k}$, and an energy that
describes the dimer oscillation such as $kb^2$. In our numerical integrations we
use a time step $\Delta t$ equal to $0.03\omega_0^{-1}$.

\begin{figure}
\includegraphics[width=10cm, clip=true]{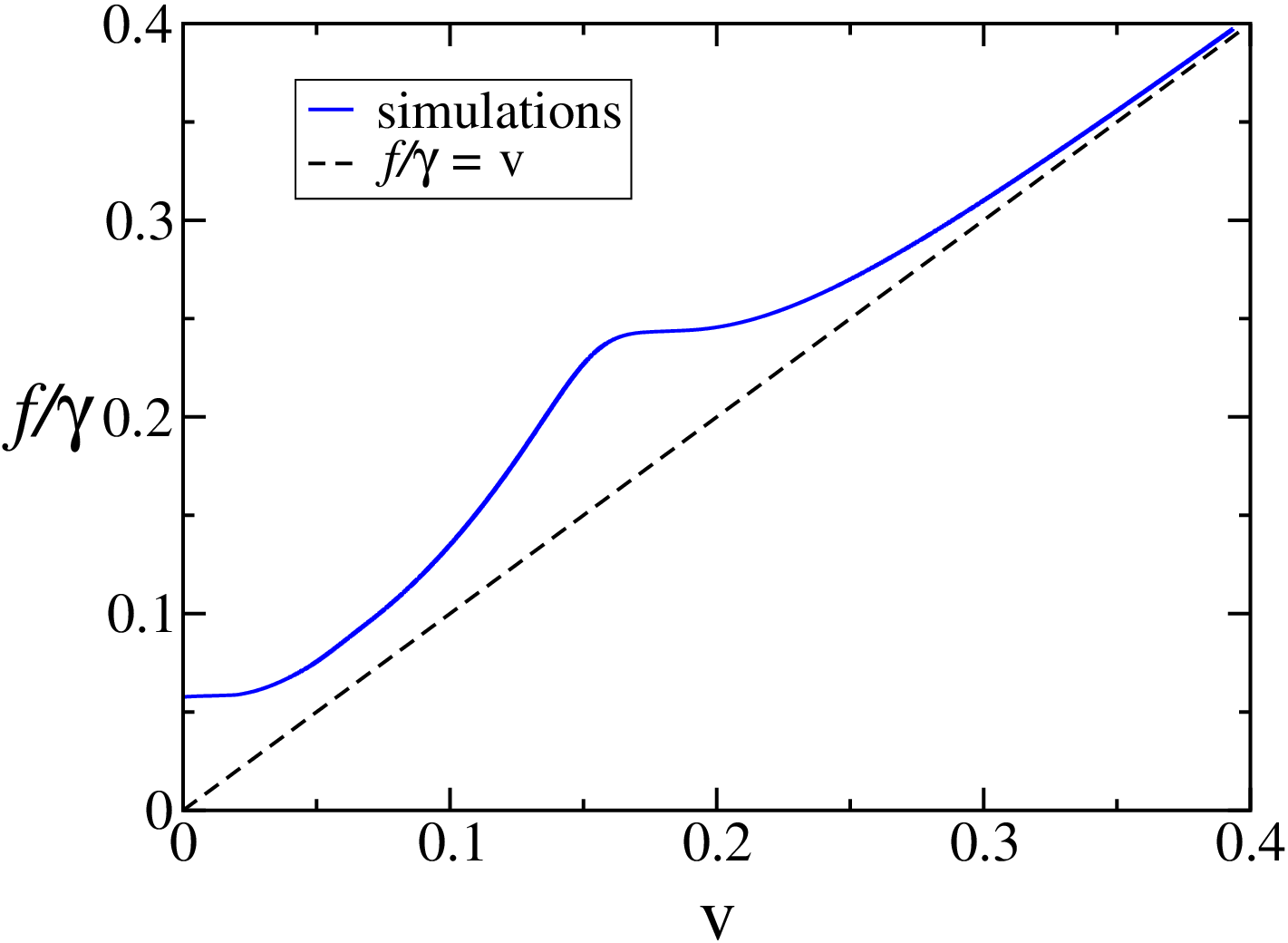}
\caption{Force--velocity ($f/\gamma$ versus $v$, $f=F/m$) relation obtained 
from numerical integration in the steady state. Both $f/\gamma$ and $v$ are
expressed in units of $b\omega_0$.  Parameter values are $u=0.038 kb^2, \gamma =
(2/3) \omega_0$, and $a/b = 1/2$.  Notice that the maximum friction at resonance
occurs when $2\pi v/b \approx \omega_0$, therefore for abscissa values which are
around $(2\pi)^{-1}$.}
\label{fxv}
\end{figure}

The procedure is as follows: for a fixed value of $F$ we obtain the center of mass
velocity $v$, averaged over several thousand time steps in the steady state.
Repeating the procedure for several hundreds of different values of
the force, we construct a characteristic curve of force versus velocity, 
$F/(m\gamma)$ versus $v$, (see Fig.\ref{fxv}).

The following features are evident from Fig.\ref{fxv} as emerging from our
numerical simulations:
\begin{enumerate}
\item
Existence of a static threshold: there is a minimum value of the externally
applied force required to make the dimer slide. It arises from the fact that, at
zero temperature, the substrate potential has to be overcome.
\item
Linear behavior in the asymptotic large-force regime: for sufficiently large
forces, the dimer slides at velocities high enough to make the substrate
potential a negligible perturbation. The linear asymptote is shown in
Fig.\ref{fxv} as the dashed line.
\item
Nonlinearity in the friction exhibiting a maximum: this arises from resonance
effects when the washboard frequency $2\pi v/b$ is close in value to the dimer
frequency $\omega_0=\sqrt{2k/m}$.
\end{enumerate}
The first two of these features are expected in light of previous reports in the
literature. The third appears to be interesting. Previous related reports have
been on the driven diffusion of a dimer~\cite{Fusco03b}, and of a
Frenkel-Kontorova model in the presence of an external force~\cite{Strunz98}.  A
thorough analysis is however lacking.  We attempt to provide such an analysis
below in the specific case of the dimer.  We relate our results to observations
made by other authors in the context of the Frenkel-Kontorova model, and suggest
that the essential features of the nonlinear friction of the infinite linear
chain can be understood in terms of the dimer dynamics we describe.

\section{Simple theoretical considerations} A transformation of the coordinates
of the two dimer masses to the center of mass coordinate $x_+=(x_2+x_1)/2$, and
the internal coordinate $x_-=(x_2-x_1)/2$, converts Eqs.(\ref{x12}) to
\begin{eqnarray}
\ddot x_+ + \gamma\dot x_+ = & &\frac{2\pi u}{m b}\sin\left(\frac{2\pi
x_+}{b}\right) \cos\left(\frac{2\pi x_-}{b}\right) + F/m \nonumber \\
\ddot x_- +
\gamma\dot x_- + \frac{k}{m}(2 x_- - a) = & & \frac{2\pi u}{m b}
\sin\left(\frac{2\pi x_-}{b}\right)\cos\left(\frac{2\pi x_+}{b}\right).
\label{x+-}
\end{eqnarray}
If in the last equation we define $\xi=\frac{2x_-}{a}-1=\frac{x_2-x_1}{a}-1$, we
get
\begin{equation}
\frac{d^{2}\xi }{dt^{2}}+\gamma \frac{d\xi }{dt}+\omega_{0}^{2}\xi =
\frac{4\pi u}{mba} \sin\left(\frac{\pi a}{b}(1+\xi)\right)
\cos \left(\frac{2\pi x_+}{b}\right)
\label{x-}
\end{equation}
as in the analysis of Ref.~\cite{Goncalves04}. Contrary to that analysis,
however, here our interest lies in the steady state reached after the
application of the external force. In that state, the
center-of-mass velocity $v_+(t)$ oscillates around a constant value
$v$. Considering only situations in which $\Delta v(t) = |v_+(t) - v| < v$, let
us neglect $\Delta v(t)$ and decouple the equations.  The internal coordinate
then satisfies
\begin{equation}
\frac{d^{2}\xi }{dt^{2}}+\gamma \frac{d\xi }{dt}+\omega_{0}^{2}\xi =
\frac{4\pi u}{mba} \sin\left(\frac{\pi a}{b}(1+\xi)\right) \cos(\omega t),
\label{xi}
\end{equation}
which describes a damped nonlinearly driven oscillator. The natural frequency is
$\omega_0 = \sqrt{2k/m}$, the damping is $\gamma$, and the driving frequency
$\omega$ is proportional to the {\em constant} component of the center of mass
velocity: $\omega = 2\pi v/b$ which is the so-called washboard frequency.

\subsection{Linear analysis in zeroth order}\label{lazo}
\begin{figure}
\includegraphics[width=10cm, clip=true]{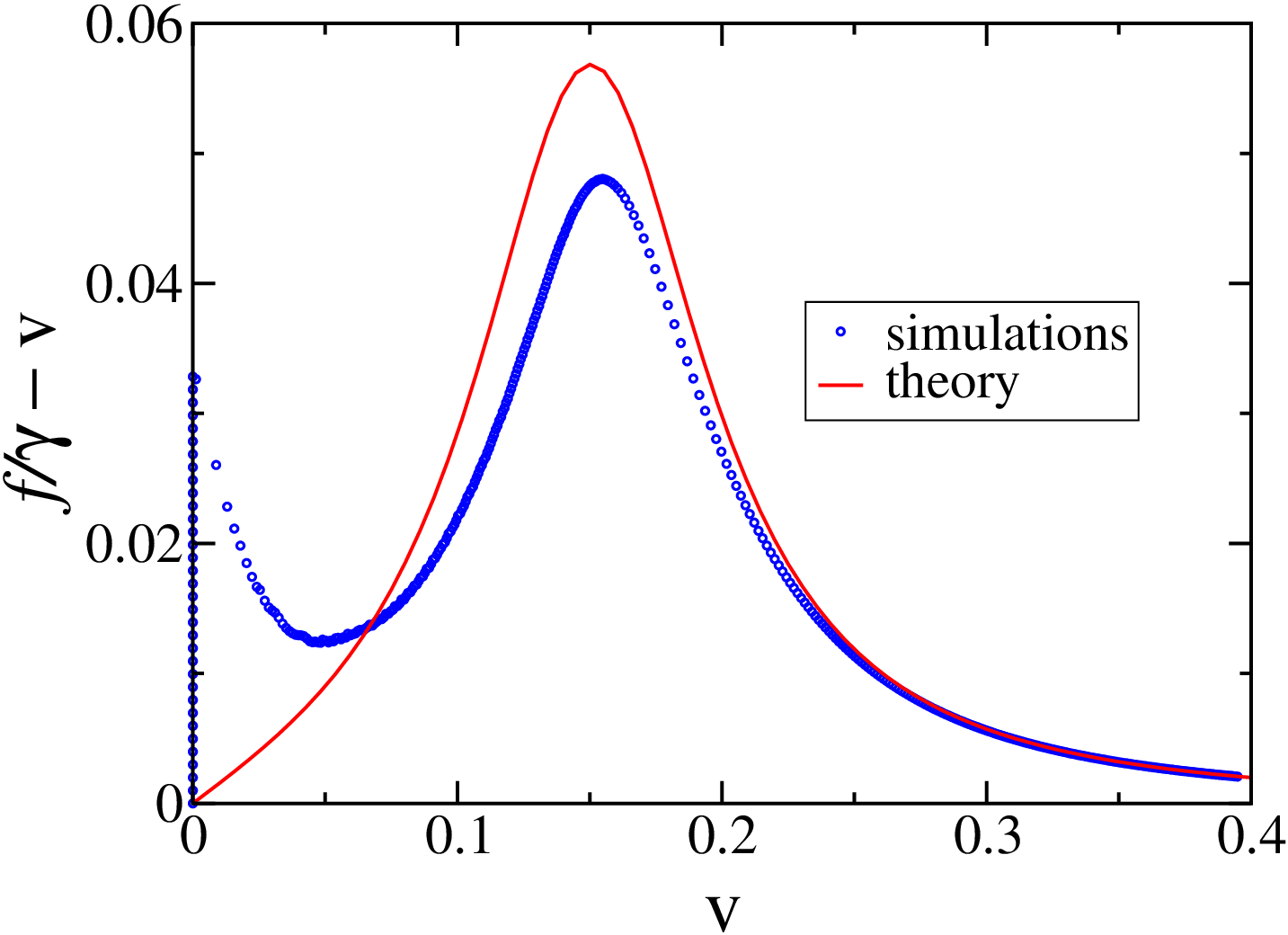}
\caption{Resonance effect in the nonlinear friction of the dimer and comparison
of the simulations with the zeroth order theory. Plotted is the nonlinear part,
$f/\gamma-v$ of the characteristic curve for a dimer sliding over a periodic
substrate. Circles are from numerical integration at the steady state. The solid
line is the zeroth order theory for the uncoupled equation for $\xi$.
$f/\gamma$ and $v$ are both in units of $b\omega_0$.  The parameters in this
case are $u=0.028kb^2, \gamma=(2/3)\omega_0$, and $a/b=1/2$.}
\label{f-v}
\end{figure}

Eq.(\ref{xi}) cannot be solved analytically because of the nonlinearity in the
sine term.  The simplest approximation, valid in zeroth order in powers of $\xi$,
leads to the equation of a driven-damped \emph{linear} oscillator:
\begin{equation}
\frac{d^{2}\xi }{dt^{2}}+\gamma \frac{d\xi }{dt}+\omega_{0}^{2}\xi =
\frac{4\pi u}{mba} \sin\left(\frac{\pi a}{b}\right)\cos(\omega t),
\label{ddo}
\end{equation}
with an exact solution in the steady state,
\begin{equation}
\xi(t) = \frac{4\pi u}{mba} \sin\left(\frac{\pi a}{b} \right)
\frac{1}{\sqrt{\left( \omega_{0}^{2}-\omega^{2}\right) ^{2}+\omega^{2}\gamma
^{2}}}\cos \left(\omega t - \delta\right),
\label{sol}
\end{equation}
where $\delta$ is the phase angle given by 
$\tan(\delta)=\frac{\gamma\omega}{\omega_0^2-\omega^2}$.
In order to compute the dependence of the friction on the center-of-mass
velocity, we must resort to the power balance condition which, in terms of the
averaged center-of-mass and internal velocity, can be written as
\begin{equation}
F<v_+> = m \gamma <v_+^2> + m \gamma <v_-^2>.
\end{equation}
Defining $f=F/m$, and using the previously made assumption that $<v_+^2> \approx
v^2$, leads to
\begin{equation}
f = \gamma v + \gamma \frac{<v_-^2>}{v}.
\label{bal}
\end{equation}
We see that, generally, the steady-state friction the center of mass of the
dimer experiences is nonlinear in the velocity.  Using the steady state solution
for $\xi(t)$ as given by (\ref{sol}), we calculate
\begin{equation}
\label{vsq}
<v_-^2> = \frac{1}{2} \left(\frac{2\pi u}{mb}\right)^2 \sin^2\left(\frac{\pi
a}{b}\right)
\frac{\omega^2}{\left(\omega_{0}^{2}-\omega^{2}\right)^{2}+\omega^{2}\gamma^{2}}.
\label{v-2}
\end{equation}
Substitution of $<v_-^2>$ given by Eq.(\ref{vsq}) in Eq.(\ref{bal}), yields
\begin{equation}
\frac{f}{\gamma} - v = \frac{1}{2}\left(\frac{u}{m}\right)^2
\sin^2\left(\frac{\pi a}{b}\right)
\frac{v}{\left(v^2 - (b\omega_{0}/2\pi)^2\right)^2 + v^2 (b\gamma/2\pi)^2}.
\label{fxv0}
\end{equation}
The right hand side of Eq.(\ref{fxv0}) focuses on the \emph{nonlinear} component
of the friction.  In Fig.\ref{f-v} we plot Eq.(\ref{fxv0}) together with the
exact results from the simulations.
We see that our zeroth order theory captures the essence of the resonance
behavior, approximates well the location of the peak, is excellent
quantitatively for larger velocities, and fails only to describe the low-velocity
threshold.

In the limit of high velocities ($v >> b\omega_0/2\pi, b\gamma/2\pi$), the
leading term in the resonance denominator is quartic in $v$, resulting in the
following nonlinear friction:
\begin{equation}
\frac{f}{\gamma} - v = \frac{1}{2} \left(\frac{u}{m}\right)^2
\sin^2\left(\frac{\pi a}{b}\right) \frac{1}{v^3}.
\label{fxvbig}
\end{equation}
This agrees with the $1/v^3$ result of the sliding friction of the purely
internally damped dimer studied recently~\cite{Goncalves04}.  In the other
limit, in the low-velocity regime ($v << b\omega_0/2\pi$), the denominator of
the right hand side of Eq.(\ref{fxv0}) is independent of $v$ (it is of fourth
order in $\omega_0$), so $f/\gamma -v$ in this case becomes linear:
\begin{equation}
\frac{f}{\gamma} - v = \frac{1}{2} \left(\frac{u}{m v_s^2}\right)^2
\sin^2\left(\frac{\pi a}{b}\right) v.
\label{fxvsmall}
\end{equation}
Here $v_s = b \omega_0 / 2 \pi$ is the sliding velocity near resonance.  Thus,
in the limit of typical velocities of experiments, like the microbalance
experiment, we recover the linear regime in which
$\frac{1}{2}\gamma\left(\frac{u}{mv_s^2}\right)^2\sin^2\left(\frac{\pi a}{b}\right)$
represents the part of the friction directly contributed by the interaction with
the substrate potential.
However, the threshold effect in the dimer
problem at $T=0$ prevents this regime from being observed.  For an extended
object and/or at $T>0$, where the threshold could be vanishing, it would be
possible in principle to observe this regime.

\subsection{Parametric oscillator in the first order} The zeroth order
description fails when $a/b$ is commensurate, i.e. when takes integer $n$
values.  In this case the zeroth order approximation cannot be applied and one
is forced to go to the next order, since neglecting $\xi$ in the sine terms
predicts an erroneous (vanishing) nonlinear friction. In such cases, and if $\xi
<< 1$, 
\begin{equation}
\sin\left(\frac{\pi a}{b}(1+\xi)\right) = (-1)^n\sin(n\pi\xi) \approx
(-1)^n n\pi\xi.
\label{sin}
\end{equation}
Substitution converts the driven-damped harmonic oscillator
equation~(\ref{xi}) into the equation for a parametric oscillator
\begin{equation}
\frac{d^{2}\xi }{dt^{2}}+\gamma \frac{d\xi }{dt}+ 
\left(\omega_{0}^{2} + (-1)^{n+1}\frac{4n\pi^2u}{mba}\cos(\omega t)\right)\xi=0.
\label{param}
\end{equation}
For this equation an exponential increase of $\xi$ is expected in an instability
window around $\omega=2\omega_0$. Thus, in this regime, $<v_-^2>$ would increase
indefinitely and the friction force would be infinite, which is of course
unphysical. In fact, in the full system Eq.(\ref{x+-}), the coupling between the
center-of-mass and internal motion drives the center of mass out of the
instability window characterizing the parametric oscillator, and this is enough
to make the increase of $\xi$ saturate~\cite{Fusco03a,Fusco03b}, yielding a 
finite value of $<v_-^2>$. Besides, if $\xi$ increased because of the
resonance, the assumption (\ref{sin}) that leads to the parametric oscillator
equation (\ref{param}) would not be valid.

\begin{figure}
\includegraphics[width=10cm, clip=true]{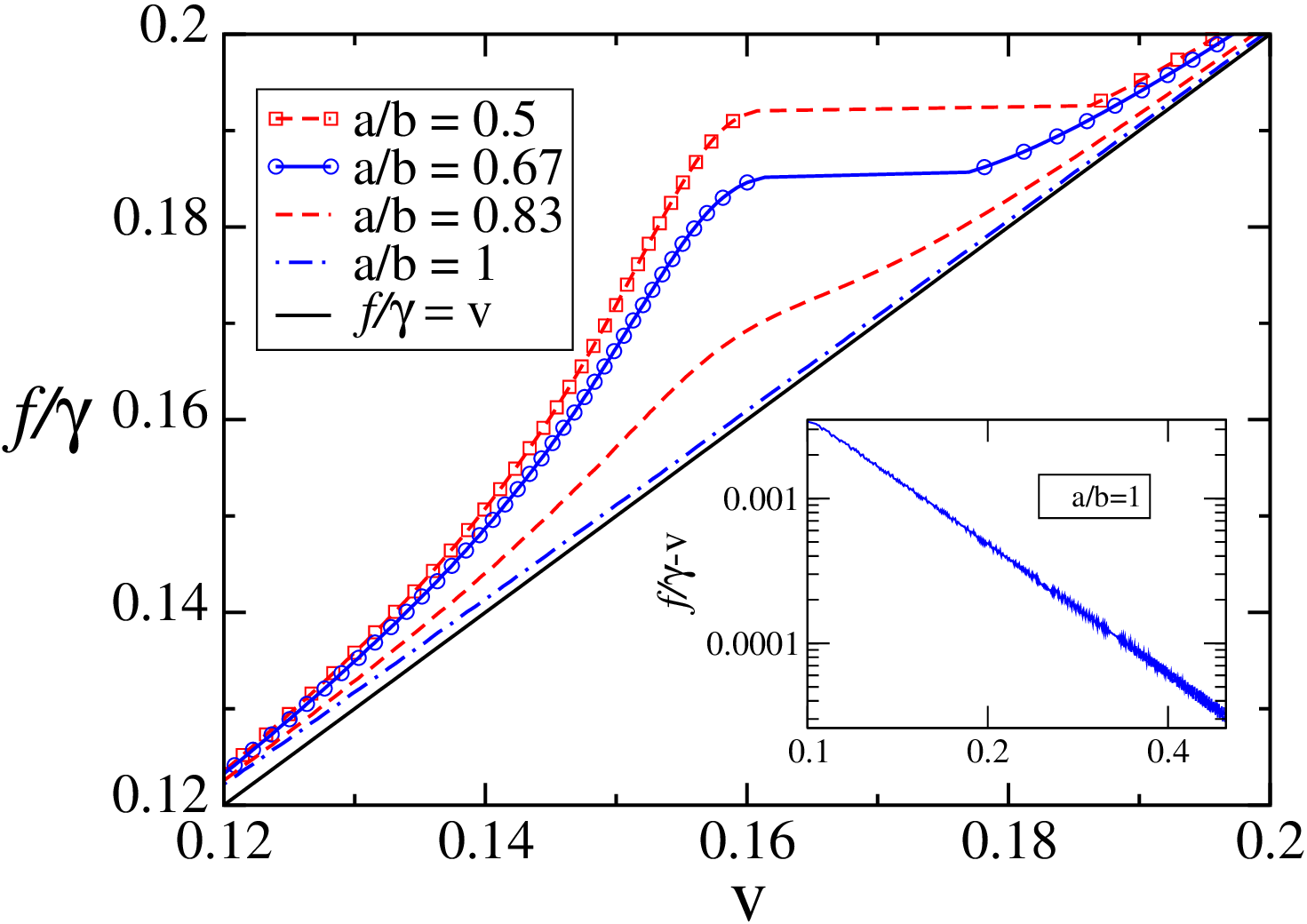}
\caption{Force--velocity relation obtained from  numerical simulation 
for different values of $a/b$.  Notice how the resonance is attenuated as $a/b$
takes on increasing values from 0.5 to 1 and how the case $a=b$ is very close
to the linear regime. The difference can only be appreciated in the inset where
$f/\gamma-v$ versus $v$ is plotted on log-log scale.  $f/\gamma$ and $v$ are both in
units of $b\omega_0$.  Parameter values are $u=0.0056kb^2,
\gamma=(1/6)\omega_0$.}
\label{fab}
\end{figure}

The zeroth order approach predicts that the shape of the resonance is proportional
to $\sin^2(\pi a/b)$. Accordingly, there are two extreme cases worth discussing:
$a/b=0.5 + n\pi$ in which the resonance is maximum, because the {\em counter
phase} movement of the particles makes this the most energy effective case; in
contrast $a/b=n\pi$ is the less effective one, since the {\em in phase} movement
does not excite the internal vibration, thus yielding purely linear
friction. The latter case is precisely the one discussed in this section.
Figure~\ref{fab} illustrates the previous discussion showing how the resonance
is actually affected by the commensuration ratio $a/b$. Those are results of
the numerical integration of Eqs.(\ref{x12}), where we can see that the zeroth
order approximation is very good indeed. In the case $a/b=1$, although the
friction is not exactly the linear one, there is no resonance at all and the
$f$--$v$ characteristic is very close to the linear one.  In the inset however,
we can see that there is nonlinear friction which goes asymptotically $v^{-3}$
to the linear regime, due to the parametric resonance.

\subsection{A non-standard description in higher order} A non-standard
approximation procedure may be developed through an iterative procedure by
noticing that, as the value of an arbitrary variable $y$ increases, the
expression $\cos(\kappa\cos(y))$, where $\kappa$ is a constant, oscillates
around the value $J_0(\kappa)$, whereas $\sin(\kappa\cos(y))$ oscillates around
$0$, $J_0$ being the Bessel function of order $0$. We have seen that, considered
in zeroth order, Eq.(\ref{xi}) predicts that $\xi(t)$ oscillates sinusoidally
with frequency
$\omega$ (see (\ref{sol})). Writing $p=\pi a/b$, we may use the fact that
\begin{equation}
<\sin\left(p(1+\xi(t))\right)> \approx <\sin(p)\cos(p\xi(t))> = \sin(p)
J_0\left(pA\sin(p)Z(\omega)\right),
\label{j0}
\end{equation}
where $A=\frac{4\pi u}{mba}$ and
$Z(\omega)=\frac{1}{\sqrt{\left(\omega_{0}^{2}-\omega^{2}\right)^{2}
+\omega^{2}\gamma^{2}}}$.  This procedure can be followed iteratively to
whatever degree is desired.  The successive approximations to the steady state
$\xi(t)$ are thus
\begin{eqnarray}
\xi^0(t) &=& A\sin(p)Z(\omega)\cos(\omega t - \delta) \nonumber \\
\xi^1(t) &=& A\sin(p)Z(\omega)J_0\left(pA\sin(p)Z(\omega)\right)
\cos (\omega t - \delta) \nonumber \\
\xi^2(t) &=& A\sin(p)Z(\omega)J_0\left(pA\sin(p)
Z(\omega)J_0\left(pA\sin(p)Z(\omega)\right)\right)\cos(\omega t - \delta) ...
\end{eqnarray}
Generally, this may be expressed by defining $\eta=A\sin(p)Z(\omega)$ and
writing the approximation as
\begin{equation}
\xi(t) \approx
\eta J_0(p\eta J_0(p\eta J_0(p\eta J_0(p\eta...))))\cos(\omega t-\delta).
\end{equation}
The spectrum predicted through this approximation is seen to be proportional to
$\omega^2\eta^2 J_0^2(p\eta J_0(p\eta J_0(p\eta J_0(p\eta...))))$ to whatever
degree of approximation one requires, except for $p \approx n\pi$, in which case
the approximation is not valid, for the same reason explained in the previous
section.

\begin{figure}
\includegraphics[width=7.5cm, clip=true]{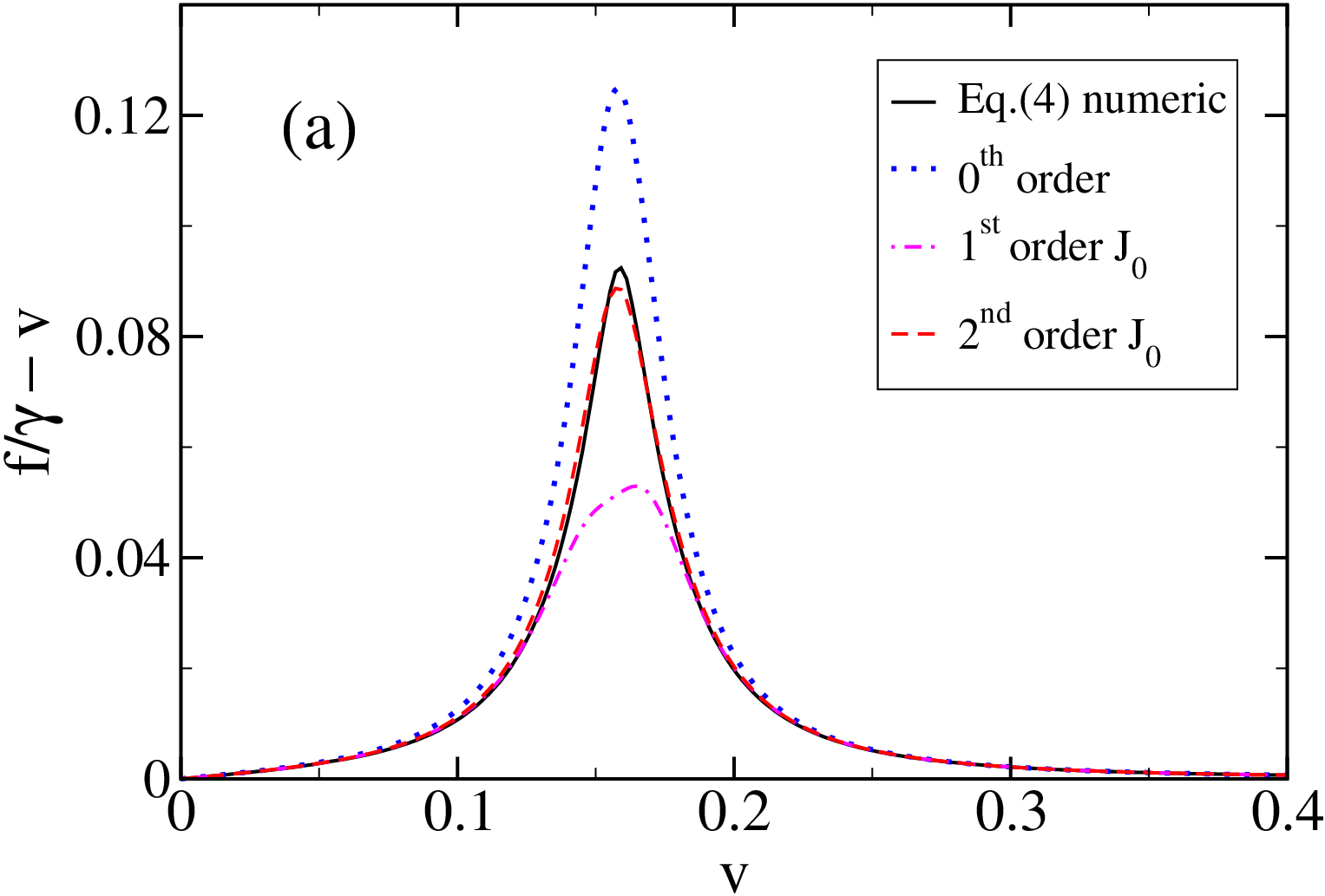}
\hspace{1cm}
\includegraphics[width=7.5cm, clip=true]{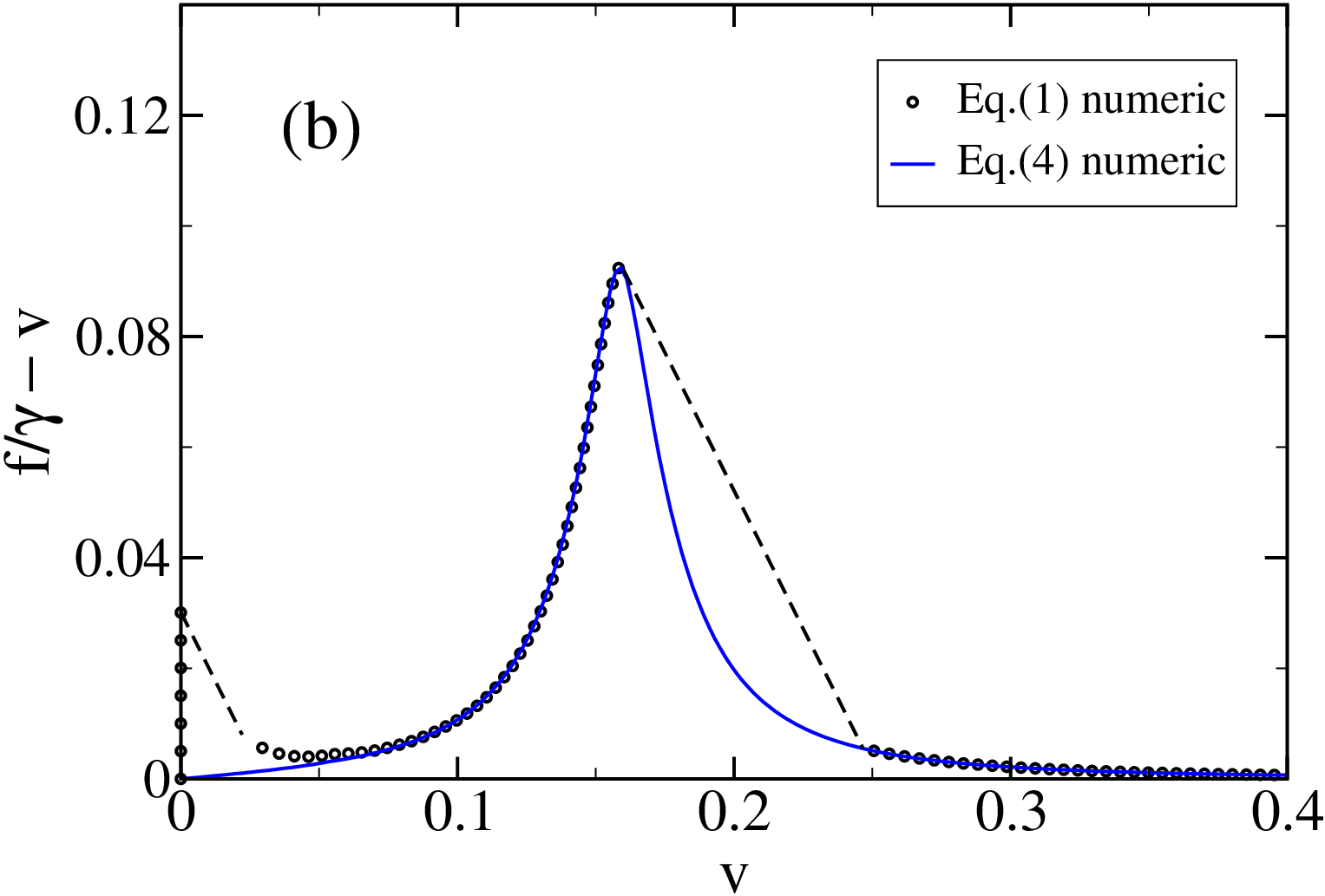}
\caption{Validity of a non-standard approximation procedure. Plotted is: 
$f/\gamma-v$ versus $v$, both in units of $b\omega_0$.
(a) Comparison between the numerical simulation of Eq.(\ref{xi}) for the 
internal coordinate $\xi$,
with the nonstandard approximation in various orders, showing excellent
convergence within the $2^{nd}$ order. (b) Comparison between exact solution of
system (\ref{x12}) with the solution of Eq.(\ref{xi}), in which the center of
mass velocity is a free parameter; notice that except for the discontinuities
the agreement is excellent.  Parameter values are $u=0.016kb^2,
\gamma=(1/4)\omega_0$, and $a/b=1/2$.}
\label{approx}
\end{figure}

The left panel of Fig.\ref{approx} shows a comparison of different orders of
this non-standard approximation with the numerical solution of Eq.(\ref{xi}),
shown as a solid line. The zeroth order approximation in this procedure, shown
as dotted line, overestimates the height of the resonance peak, while the first
order approximation, shown as dashed-dotted line, underestimates it. In the
second order, shown as dashed line, our procedure is already able to essentially
coincide completely with the numerical solution.  It is important to realize
that our non-standard procedure does very well within the second order when
viewed as an approximation to Eq.(\ref{xi}), for which it has been developed,
rather than to the original Eqs.(\ref{x12}).  The right panel shows the
comparison of the numerical solution of Eq.(\ref{xi}), in which the center of
mass velocity is a constant, with the simulation based on the original
Eqs.(\ref{x12}). One can see a low-velocity departure arising from the static
threshold and a high-velocity departure arising from bistability. Such differences
arise from the fact that in Eqs.(\ref{x12}) the center-of-mass velocity is
itself decided by the dynamics, therefore not being a free parameter as in
Eq.(\ref{xi}).

\section{Further Nonlinear Results}
\subsection{Bistability and Hysteresis}
\begin{figure}
\includegraphics[width=7.5cm, clip=true]{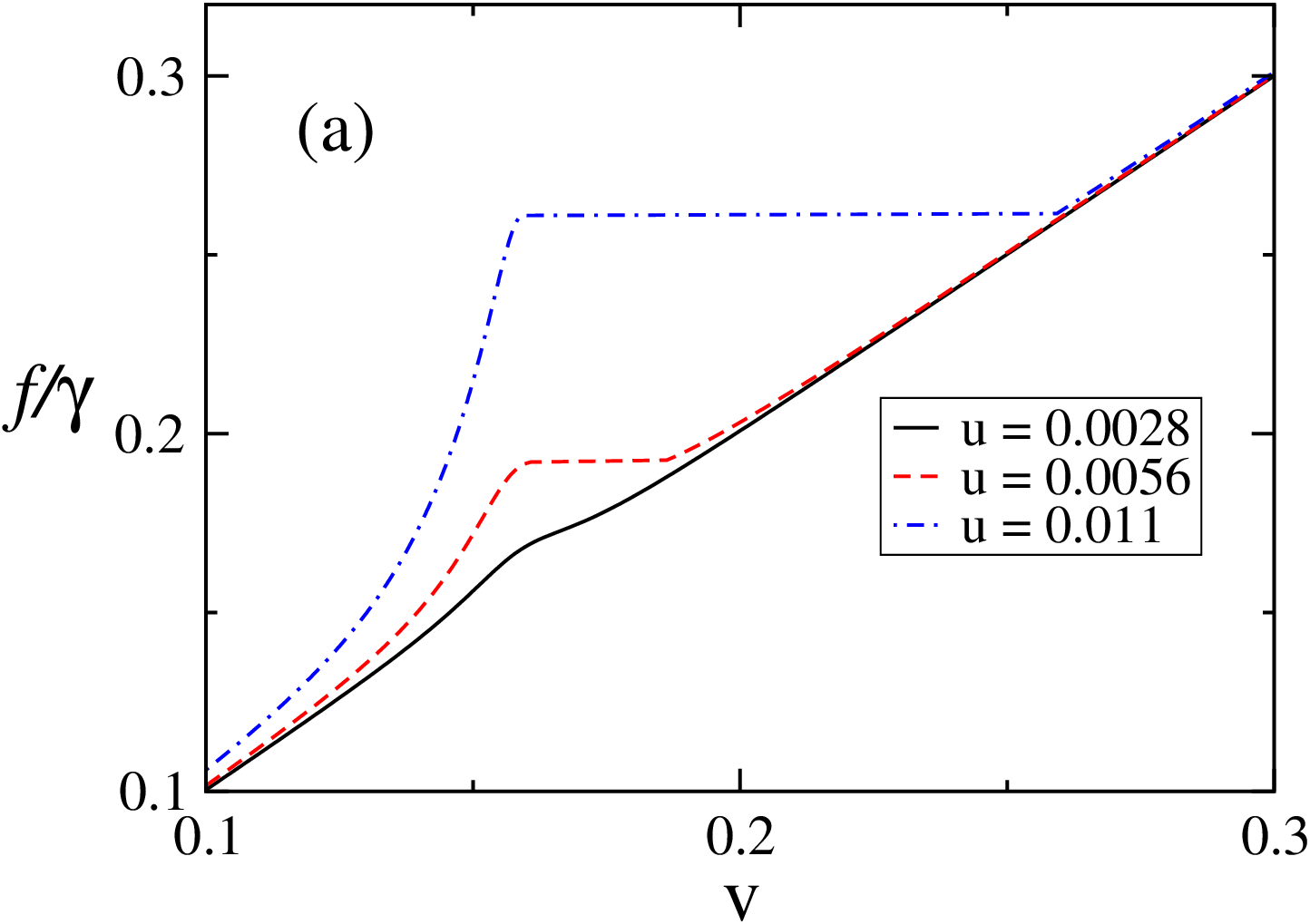}
\hspace{1cm}
\includegraphics[width=7.5cm, clip=true]{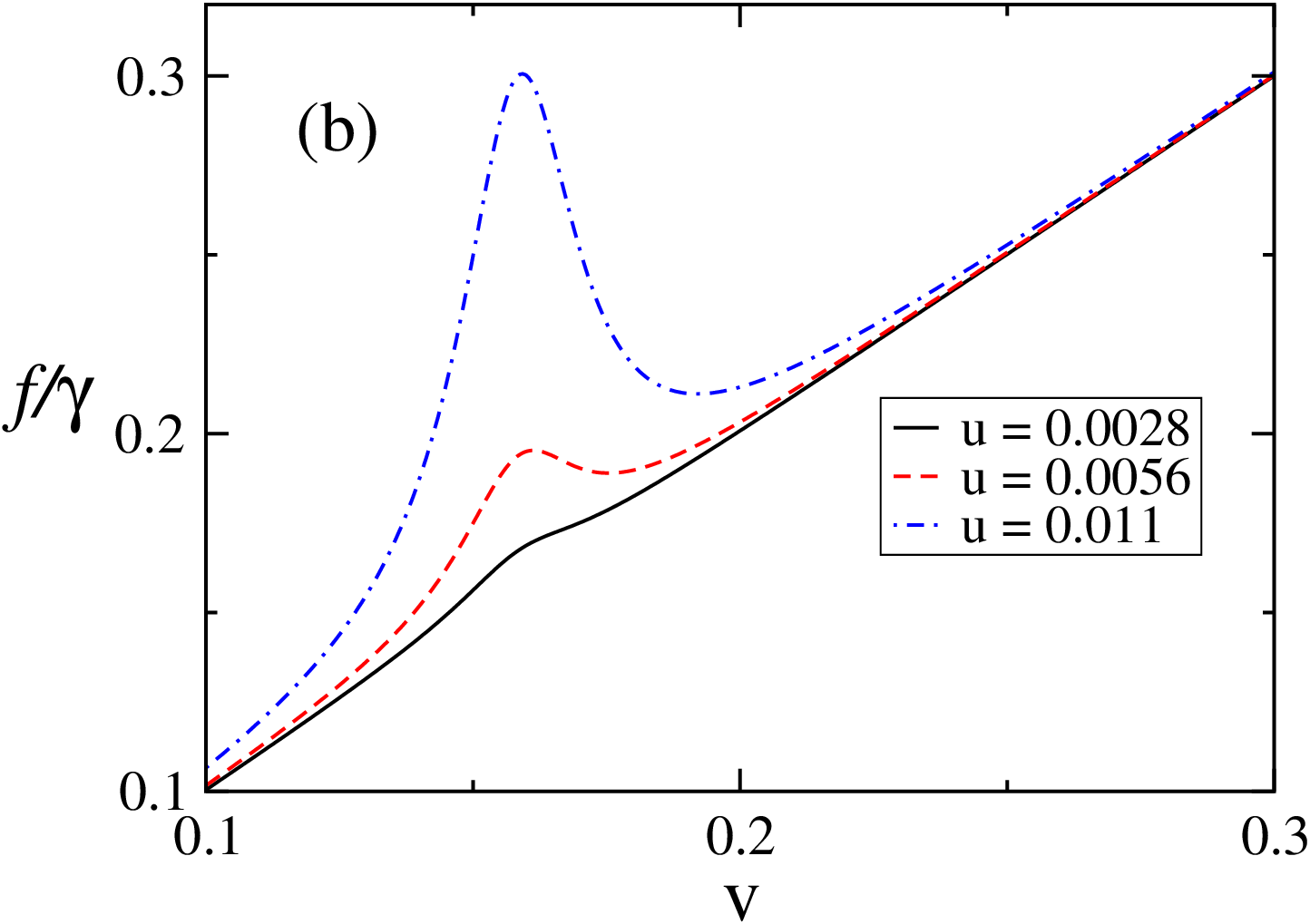}
\caption{Force--velocity curves $f/\gamma$ versus $v$ (both in units of $b\omega_0$)
 for different values of the corrugation amplitude $u$ expressed in units of
 $kb^2$.  Parameter values are: $a/b=0.5, \gamma=(1/6)\omega_0$. (a) Simulation
 results (numerical integration of Eqs.(\ref{x12})). (b) Zeroth order
 approximation (Eq.(\ref{fxv0})).  Notice in both panels that the critical value
 of $u$ (i.e. the value at which the bistability develops) is around $0.004kb^2$
 for the chosen values of parameters.}
\label{fxu}
\end{figure}

\begin{figure}
\includegraphics[width=10cm, clip=true]{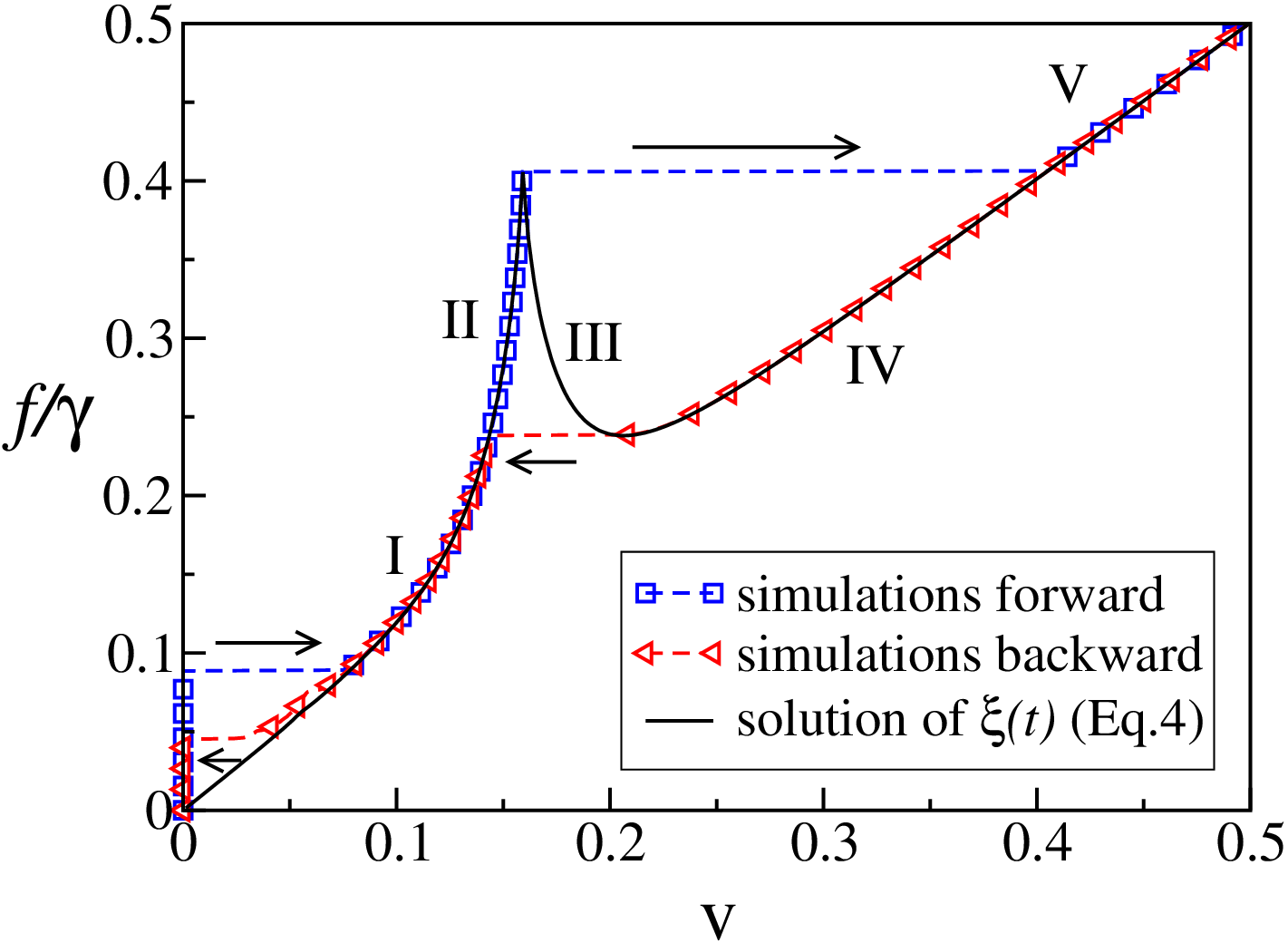}
\caption{Hysteresis in the force--velocity relation. Symbols are results from
simulations of the system Eqs.(\ref{x12}): ($\square$) increasing the
external force, ($\lhd$) decreasing the external force.
The solid line is the result of numerical integration of the Eq.(\ref{xi}).
I and V indicate the stable regions (overlap of the two symbols), II and IV are
metastable regions, while III denotes the unstable region. The arrows indicate
the forward jumps ($\rightarrow$) and backward jumps ($\leftarrow$).
Plotted is  $f/\gamma$ versus $v$ (expressed in units of $b\omega_0$). Parameter
values are $u=0.023 kb^2, \gamma= (1/6)\omega_0$, and $a/b = 1/2$.}
\label{hyst}
\end{figure}

The nonlinearities present in our system give rise to bistability.  This can be
seen in Fig.\ref{fxu} where it is clear that one value of the force can
correspond to two distinct values of the velocity, in a certain region.  In
order to gain deeper insight into this issue, we plot in Fig.\ref{fxu}(b) the
prediction for the characteristic curve based on expression~(\ref{fxv0}) as $u$
increases. We see features typical of bistable systems such as encountered in
the pressure-volume ($p$--$V$) diagram of a van der Waals gas.
The bistability of the dimer is directly related to hysteresis.  Fig.\ref{hyst}
shows the presence of two regions, identified by the dashed lines, where
hysteresis occurs when the force is first increased from $f=0$ and then is
decreased in small steps down to $f=0$. The first hysteresis at low velocities
is due to the bistability between the locked and the running state and has a
static origin, being associated with the energy threshold that the particle has
to overcome in order to move. The same kind of hysteresis is also found in the
underdamped monomer~\cite{Risken}.  On the other hand, the second hysteresis at
intermediate values of $v$ is of purely dynamical nature, and it is related to
the bistability between two running states in that region. Let us discuss the
latter hysteresis in more detail. Five regions, denoted by roman numbers, can be
identified there : I) the mechanically stable region where $v$ increases with
$F$ up to a point where the force begins to be bivaluated, II) the metastable
region where $v$ increases with $F$ up to the local maximum, III) the
mechanically unstable region where $v$ decreases with $F$ up to the local
minimum, IV) the metastable region, where $v$ increases with $F$ from the local
minimum up to a point where the force ends to be bivaluated, and V) the
mechanically stable region where $v$ increases again with $F$ and it is
monovaluated. Region I) and V) are both obtained either increasing the force
from zero or decreasing it from a high value. Region II, however is accessible
while increasing the force from zero but not when the force is decreased from a
high value. Complementary, region IV is obtained while decreasing the force from
a high value but not when the force is increased from zero.  Region III) is
mechanically unstable, i.e the velocity decreases when the external force is
increased. For a given value of the external force, the system goes to a stable
point, therefore when the force goes beyond the value at the local maximum there
is a jump from region II to region V.  Conversely when the force is decreased
from region V the system enters the metastable region IV until at the local
minimum it jumps to region I.  Notice that, except for the unstable region III,
all the hysteresis features of the system are excellently reproduced by the
solution of Eq.(\ref{xi}).

\begin{figure}
\includegraphics[width=10cm, clip=true]{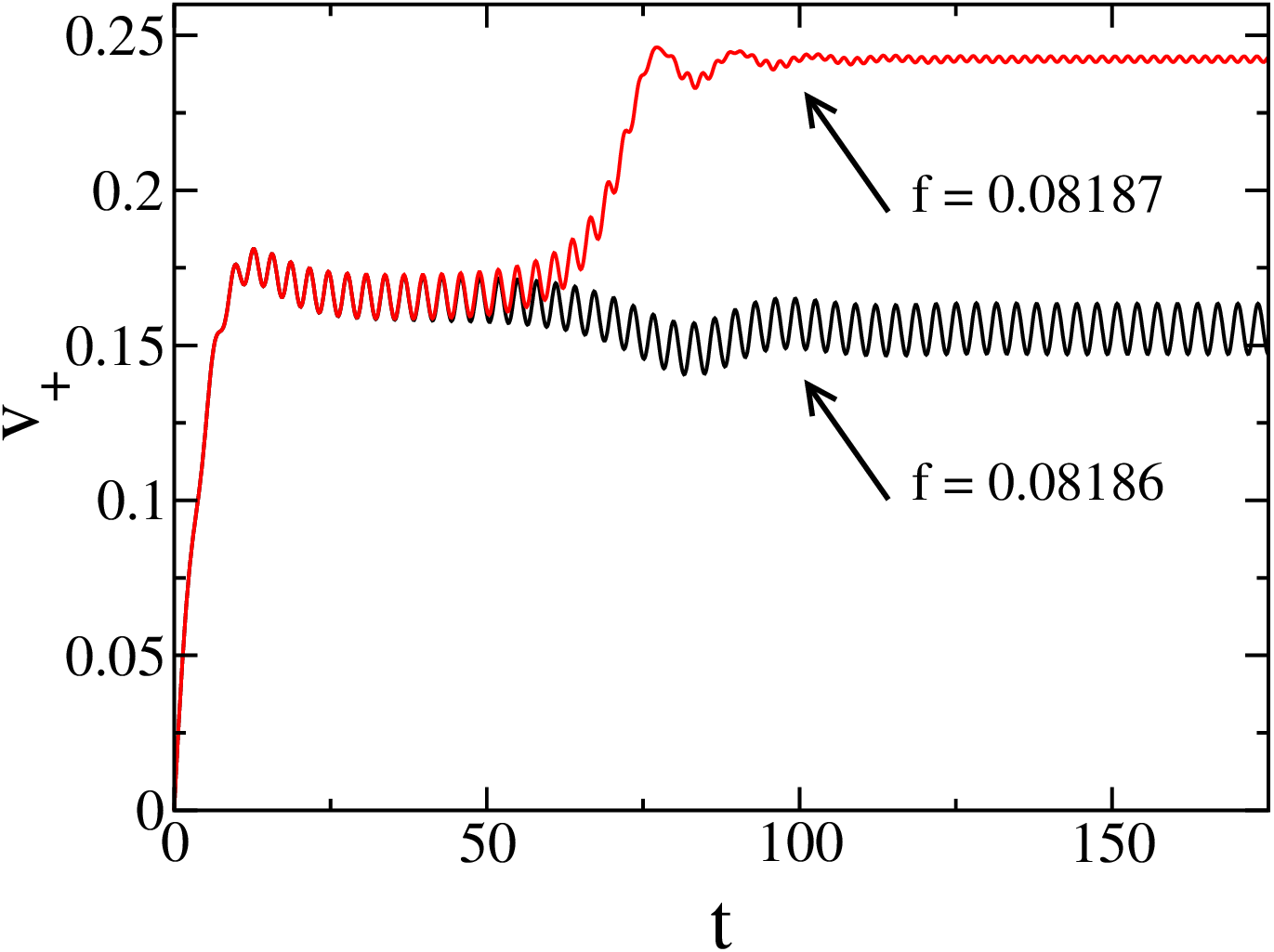}
\caption{Bistable behavior exhibited in the time evolution of the center-of-mass
velocity for slightly different values of the external force $f=F/m$. Velocity
$v_+$ is in units of $b\omega_0$, time $t$ in units of $\omega_0^{-1}$, and $f$
in units of $kb$. Notice that, up to time $t=60\omega_0^{-1}$, the two curves follow
almost the same dynamics, but then separate, respectively, to a low and high
limit. Parameters are $u=0.011 kb^2,
\gamma = (1/6)\omega_0$, and $a/b = 1/2$.}
\label{vel}
\end{figure}

Fig.\ref{vel} shows the behavior of the center-of-mass velocity for two
different, but close, values of the external force near the region of
bistability. In the early stages of the dynamics the velocity is practically the
same in the two cases, but after some time the velocity corresponding to the
lower force attains a steady state value that is much smaller than the one
corresponding to the larger force. This clearly illustrates the dynamical origin
of the bistability.  Furthermore, the oscillations of $v_{+}$ in the steady
state are much smaller for the larger force.  The bistable behavior of the dimer
critically depends on the parameters $u$, $\omega_0$ and $\gamma$: For fixed
$\omega_0$ and $\gamma$, it is observed when $u$ exceeds a critical value, which
can be estimated in the framework of the zeroth order approximation as the value
for which the local maximum and the local minimum in the velocity-force
characteristic, given by Eq.(\ref{fxv0}) by adding the linear term in $v$,
coincide. This is related, in the van der Waals analogy, to the region of
coexisting phases.  The interplay between linear and nonlinear friction due to
resonance gives rise to the bistability, as does the interplay between
attractive and repulsive terms in the van der Waals gas.  In this sense,
Eq.(\ref{fxv0}) might be regarded as an equation of state of the system.  Both
equations are however only approximate descriptions of the real system and
cannot predict what happens in the transition region, where a coexistence of two
different states is found.

Since the hysteresis is intimately linked to the bistability, it appears only
for large $u$ (and/or small $\gamma$), as discussed above.

\subsection{Effects of non-zero temperature}
\begin{figure}
\includegraphics[width=7.5cm, clip=true]{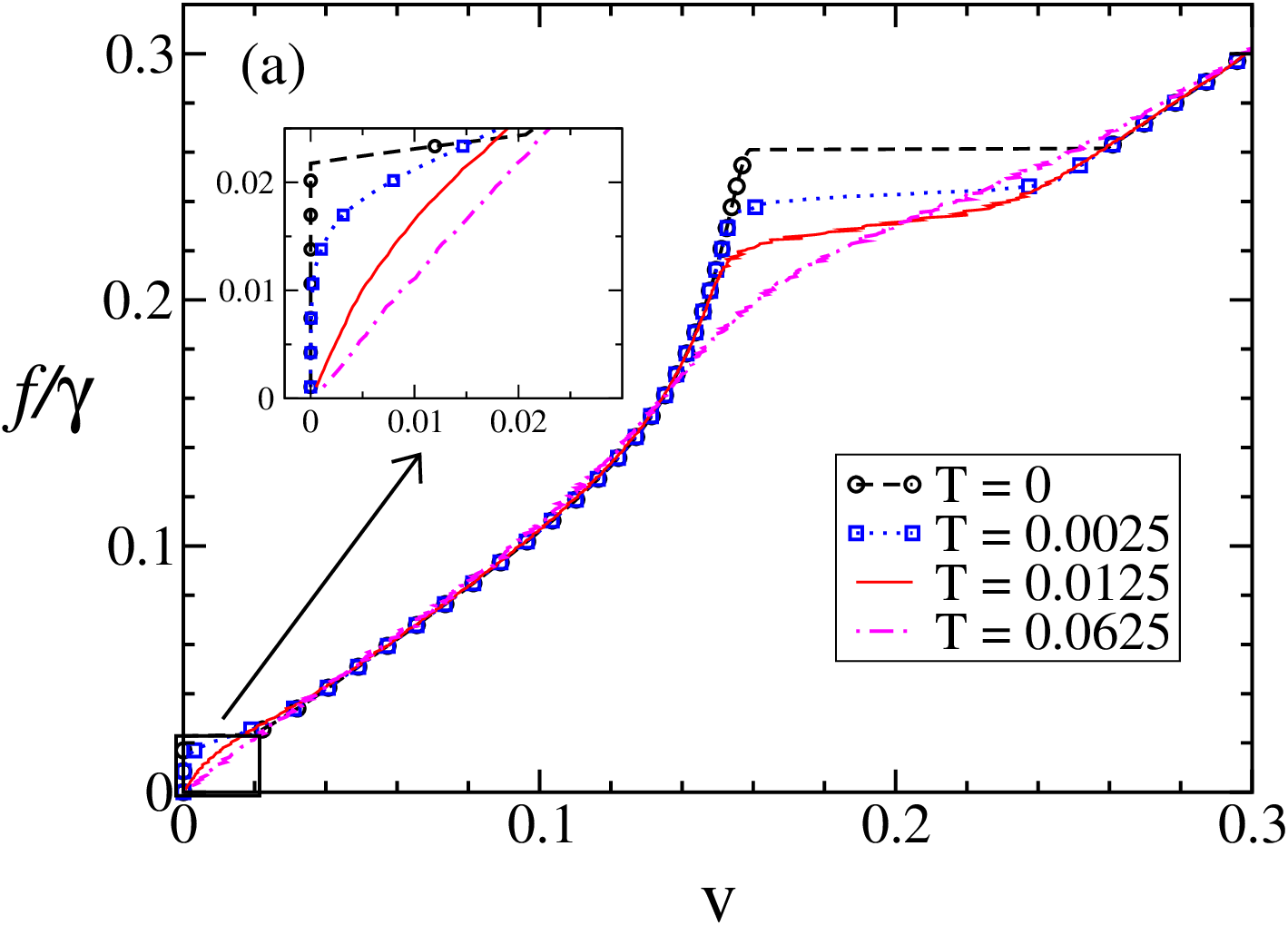}
\hspace{1cm}
\includegraphics[width=7.5cm, clip=true]{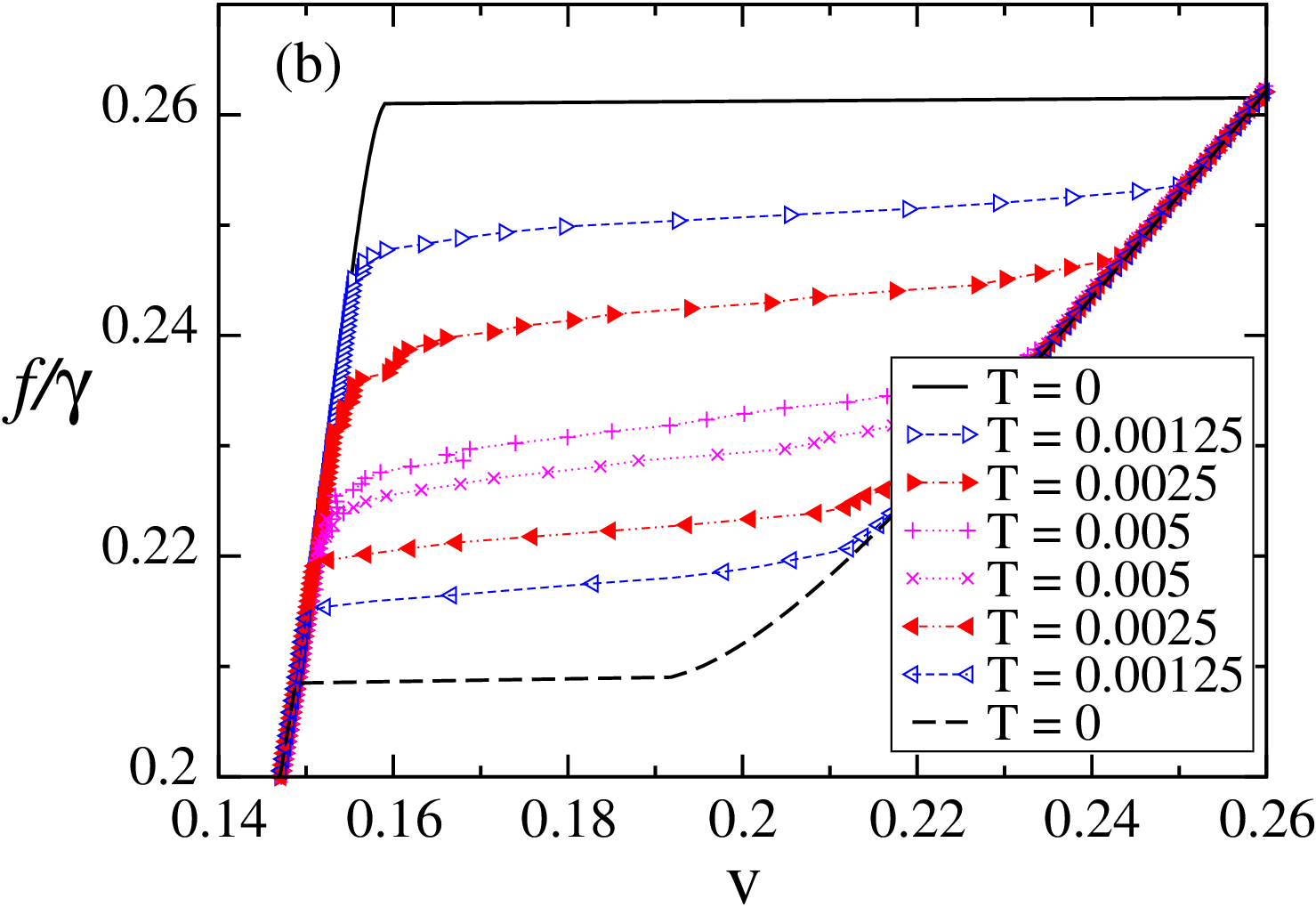}
\caption{Temperature effects on the force--velocity relation. Plotted is the 
$f/\gamma$ versus $v$ relation as temperature $T$ (expressed in units of
$u/k_B$) is varied.  $f/\gamma$ and $v$ are in units of $b\omega_0$, and
parameters are $u=0.011 kb^2, \gamma = (1/6)\omega_0$, and $a/b = 1/2$.  (a)
Curves for increasing force, the inset shows the details in the static threshold
region. The bistability is seen to disappear as $T$ increases smoothening the
characteristic $f$--$v$ curve.  (b) Forward and backward curves to emphasize the
hysteresis dependence on temperature: the area of the hysteresis loop decreases
with $T$ and disappears for temperatures close to $k_B T=0.005u$.}
\label{temp}
\end{figure}

It is natural to inquire into the effects of finite temperature in our system.
We solve Eqs.(\ref{x12}) by adding random forces representing the thermal
interaction with the substrate, in the Langevin approach:
\begin{eqnarray}
m\ddot x_1 + m\gamma\dot x_1 + k(x_2 - x_1 - a) =
\frac{2\pi u}{b}\sin\left(\frac{2\pi x_1}{b}\right) + R_1 + F \nonumber \\
m\ddot x_2 + m\gamma\dot x_2 - k(x_2 - x_1 - a) =
\frac{2\pi u}{b}\sin\left(\frac{2\pi x_2}{b}\right) + R_2 + F,
\label{lan}
\end{eqnarray}
in which the stochastic forces $R_{1,2}$ satisfy the conditions
\begin{eqnarray}
<R(t)> & = & 0 \nonumber \\ <R(t)R(t^\prime)> & = & 2\gamma mk_BT\delta (t-t^\prime),
\end{eqnarray}
where $k_B$ is Boltzmann's constant and $T$ is temperature.  Fig.\ref{temp}
illustrates the velocity-force characteristics at finite temperature, for
different values of $T$. By increasing $T$, the effects we have shown at $T=0$
are increasingly smeared out. In particular, the static threshold disappears, as
can be seen in the inset of the left panel of Fig.\ref{temp}. The bistability
regions and hysteresis still survive up to small values of $T$, and the
characteristic curve is smoothened in the region of dynamical bistability.
Interestingly, the area of the hysteresis loop, shown in the right panel of
Fig.\ref{temp}, decreases with $T$ and eventually disappears for sufficiently
high temperatures. A hysteretic behavior in the intermediate friction region, at
$T\ne 0$, has been reported for long periodic chains~\cite{Braun97}. Here we see
that we recover essentially the same behavior in the case of the dimer. This is
because the same mechanisms are effective, as will be discussed in the next
section.

\begin{figure}
\includegraphics[width=7.04cm, clip=true]{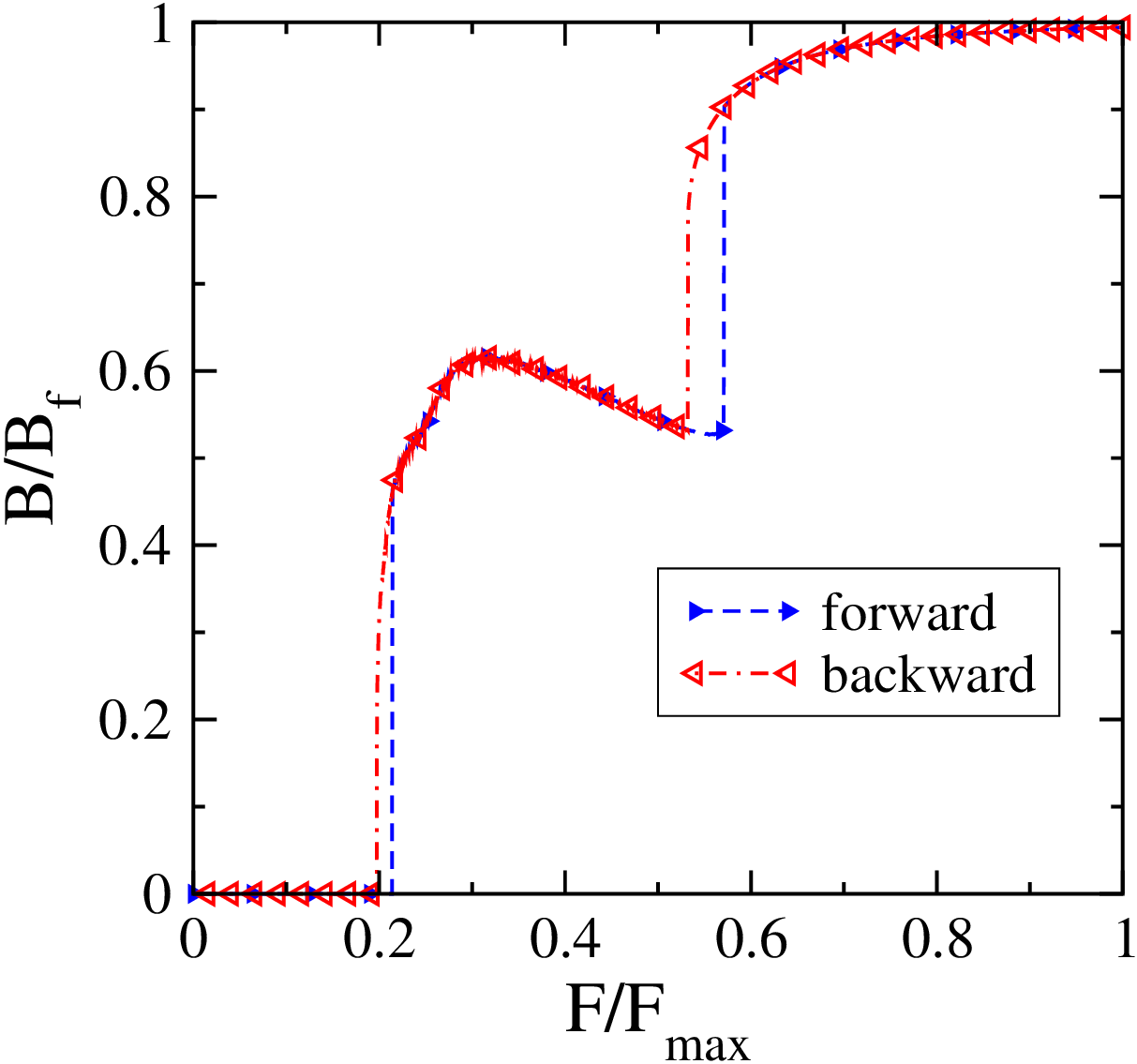}
\caption{Mobility $B=<v>/F$, normalized by the asymptotic mobility at
large forces $B_f=(m\gamma)^{-1}$, as a function of the driven force $F$. This
plot should be compared to that given for the Frenkel-Kontorova model by Braun
et al.~\cite{Braun97} (see their Fig.1a).  The force is normalized by the
maximum applied value $F_{max}=0.71kb$.  The parameters in this case are $u=0.056
kb^2, \gamma = (2/3)\omega_0$, and $a/b = 1/2$.}
\label{mobility}
\end{figure}

\section{concluding remarks} 
Our general aim in the present paper has been to extend previous
investigations~\cite{Goncalves04} of friction in the simplest non-trivial
system, a dimer moving over a periodic substrate, to include a driving force and
arbitrary temperatures.  Our study has shed some light on puzzling features in
the literature for more complex systems, such as the Frenkel-Kontorova chain, by
allowing us to obtain simple physical arguments for our system.  The model is
indeed simple: a linear damped oscillator sliding in a sinusoidal periodic
potential. Yet, except for the limitation that it is restricted to a single
spatial dimension, it has the necessary ingredients to represent a real dimer or
molecule set in a controlled microscopic sliding experiment.  For example, a
molecule sliding along channels of a crystalline well-oriented
substrate~\cite{Kurpick} should exhibit some of the features we have described.
Our results, both analytical and numerical, confirm the existence of nonlinear
friction, resonance effects, bistability, and hysteresis, which can be well
understood in terms of the resonance of a driven, damped oscillator. Far from
resonance the sliding friction goes asymptotically to the linear regime with a
$v^{-3}$ term, which represents the tail of the resonance response. For
intermediate forces, bistability and hysteresis emerge as a consequence of the
interplay between linear friction and resonance. That our results provide a
simple representation of phenomena reported for large systems should be clear
from Fig.\ref{mobility}, which shows a comparison between our results on the dimer
and results from Braun et al.~\cite{Braun97} for the Frenkel-Kontorova model,
which can be regarded as the infinite-size generalization of the dimer. Those
results are presented to facilitate direct comparison with Fig.1a of
Ref~\cite{Braun97} in terms of the mobility $B=v/F$, normalized by its
asymptotic value $B_f=1/m\gamma$.  We notice that the intermediate behavior
before the asymptotic linear regime is observed in the same fashion as for the
dimer.  In the latter case the mechanism is completely understood as being due
to resonance of the dimer; the same mechanism could also underlie the nonlinear
friction in the Frenkel-Kontorova case (see Ref.~\cite{Strunz98}).  Braun at
al.~\cite{Braun97} attributed such features to the presence of kinks that can be
observed during the sliding regime of the Frenkel-Kontorova chain.

In light of the present study, let us make some comments that might have some
relevance to the ongoing debate in the
literature~\cite{Smith,Persson96,Tomassone97,Liebsch99} about whether
sliding friction is mainly electronic or phononic in origin. For this purpose
let us identify the background (linear) friction in our model with electronic
and the resonance friction with phononic sources. Such an identification appears
natural because, although vibrations of the substrate do not exist in our model,
resonance friction does arise from the interaction of the dimer vibrations with
the substrate; on the other hand, the source of the intrinsic friction, denoted
in our analysis by $\gamma v$, is independent of interactions with the
substrate. We have seen from our analysis that the resonance friction is
modulated by $sin^2(\pi a/b)$. The commensuration ratio $a/b$ appearing in this
modulation factor would depend in a realistic 2-d or 3-d environment
additionally on the relative orientation of the dimer (adlayer) and the
substrate. Therefore, while resonance friction might dominate the background
friction in principle (for sufficiently large corrugation amplitude values $u$),
the smallness of the modulation ratio could make it have disparately small
values relative to background friction. That might be a plausible explanation
for the disparate results obtained in otherwise similar simulation
models~\cite{Smith,Persson96,Tomassone97,Liebsch99}.  However, rather high
velocities seem to be necessary for resonance friction to be appreciable. Rough
estimates we have made suggest that in a number of materials, $b\approx 2$\AA,
$\omega_0 \approx 10-100$ cm$^{-1}$, adlayer velocities relative to the
substrate necessary for resonance friction to be observable would be as high as
$30-300$ m/s. The velocity region of the quartz microbalance experiments
corresponds to the low velocity limit discussed at the end of section \ref{lazo}
(see Eq.\ref{fxvsmall}) where the resonance
friction becomes linear and the ratio between {\em phononic} and {\em electronic} 
friction results proportional to $\left(\frac{u}{k^2b^2}\right)^2$. 
Let us extend this estimation to the case of
a Lennard-Jones model for the dimer (adsorbate) interaction, where that ratio
becomes proportional to $(u/\epsilon)^2$, $\epsilon$ being the well depth of the
Lennard-Jones potential. In the Xe over Ag case for example $\epsilon \approx
20$meV and $u$ can lie in the 1-2meV range, therefore the prefactor
$(u/\epsilon)^2 \approx 0.01$. With the above suggested identification of the
friction mechanisms ---and within the validity of analytical assumption used in
the low velocity limit---, one might thus expect phononic friction to be rarely
observable for most materials under typical experimental conditions.

\section{acknowledgments}
This work was supported in part by a grant made by the Los Alamos National
Laboratory to the Consortium of the Americas of the University of New Mexico,
and by the National Science Foundation under grants INT-0336343 and
DMR-0097204. S.G. and C.F. acknowledge the hospitality of the Department of
Physics and Astronomy of the University of New Mexico, during their stay at the
Consortium of the Americas. Work at Los Alamos is supported by the USDOE.

\section{references}

\end{document}